\documentclass[conference]{IEEEtran}
\IEEEoverridecommandlockouts
\usepackage{cite}
\usepackage{amsmath,amssymb,amsfonts}
\usepackage{algorithmic}
\usepackage{graphicx}
\usepackage{textcomp}
\usepackage{xcolor}
\def\BibTeX{{\rm B\kern-.05em{\sc i\kern-.025em b}\kern-.08em
    T\kern-.1667em\lower.7ex\hbox{E}\kern-.125emX}}
\begin{document}

\title{
Cooperative Beamforming and RISs Association for
\\ Multi-RISs Aided Multi-Users MmWave \\
MIMO Systems through Graph Neural Networks

}

\author{\IEEEauthorblockN{Mengbing Liu\textsuperscript{1}, Chongwen Huang\textsuperscript{2}, Marco Di Renzo\textsuperscript{3}, M\'{e}rouane Debbah\textsuperscript{4} and Chau Yuen\textsuperscript{1}}
\IEEEauthorblockA{{\textsuperscript{1}Engineering Product Development Pillar, Singapore University of Technology and Design, Singapore} \\
{\textsuperscript{2}College of Information Science and Electronic Engineering, Zhejiang University, China}\\
{\textsuperscript{3}Universit\'{e} Paris-Saclay, CNRS, CentraleSup\'{e}lec, Laboratoire des Signaux et Syst\`{e}mes, Gif-sur-Yvette, France} \\
{\textsuperscript{4}Technology Innovation Institute, Masdar City, Abu Dhabi, United Arab Emirates}\\
Email: mengbingliu714@gmail.com; chongwenhuang@zju.edu.cn; marco.di-renzo@universite-paris-saclay.fr;\\ merouane.debbah@tii.ae; yuenchau@sutd.edu.sg
}
\thanks{The work of Prof. Huang was supported by the China National Key R$\&$D Program under Grant 2021YFA1000500, National Natural Science Foundation of China under Grant 62101492, Zhejiang Provincial Natural Science Foundation of China under Grant LR22F010002, National Natural Science Fund for Excellent Young Scientists Fund Program (Overseas), Zhejiang University Education Foundation Qizhen Scholar Foundation, and Fundamental Research Funds for the Central Universities under Grant 2021FZZX001-21.}
\thanks{This research is supported by the Ministry of Education, Singapore, under its MOE Tier 2 (Award number MOE-T2EP50220-0019). Any opinions, findings and conclusions or recommendations expressed in this material are those of the author(s) and do not reflect the views of the Ministry of Education, Singapore. }}

\maketitle

\begin{abstract}
Reconfigurable intelligent surface (RIS) is considered as a promising solution for next-generation wireless communication networks due to a variety of merits, e.g., customizing the communication environment. Therefore, deploying multiple RISs helps overcome severe signal blocking between the base station (BS) and users, which is also a practical and effective solution to achieve better service coverage. However, reaping the full benefits of a multi-RISs aided communication system requires solving a non-convex, infinite-dimensional optimization problem, which motivates the use of learning-based methods to configure the optimal policy. This paper adopts a novel heterogeneous graph neural network (GNN) to effectively exploit the graph topology in the wireless communication optimization problem. First, we characterize all communication link features and interference relations in our system with a heterogeneous graph structure. Then, we endeavor to maximize the weighted sum rate (WSR) of all users by jointly optimizing the active beamforming at the BS, the passive beamforming vector of the RIS elements, as well as the RISs association strategy. Unlike most existing work, we consider a more general scenario where the cascaded link for each user is not fixed but dynamically selected by maximizing the WSR. Simulation results show that our proposed heterogeneous GNNs perform about 10 times better than other benchmarks, and
a suitable RISs association strategy is also validated to be effective in improving the quality services of users by 30$\%$.
\end{abstract}


\section{Introduction}

Multiple-input multiple-output (MIMO) technology continues to play an indispensable role in providing high spectral efficiency in the current and next-generation wireless systems \cite{bjornson2019massive}. Although MIMO technology can bring significant antenna gains, the signal transmitted at millimeter wave (mmWave) bands experiences serious propagation attenuation.
%, which is vulnerable and easily blocked. 
Moreover, due to the existence of obstacles in the environment, MIMO technology cannot provide seamless coverage between transceivers \cite{xiao2017millimeter}.  As a remedy, reconfigurable intelligent surface (RIS) is considered as a promising technology to improve the links' quality of service as well as recover communications in the dead zone. Compared with other technologies, such as ultra dense networks, RIS is regarded as a promising paradigm to assist mmWave MIMO networks via a large number of effective and low-cost reflective elements  \cite{huang2019reconfigurable,liu2022deep,huang2020reconfigurable}. 

Due to the benefits above,  RIS has been introduced to various communication systems to balance the network capacity and energy consumption. To fully reap the benefits, the cooperative beamforming design on base stations (BSs) with MIMO and RISs is a crucial challenge since many beamforming designs for the distributed RISs system are non-convex and heavily computational. Thus, a great deal of efforts have been devoted to developing effective algorithms to tackle this challenge in RIS-aided systems \cite{ning2020beamforming,cao2020intelligent}. Recently, the cooperative beamforming design in multi-RISs aided system has been investigated to compensate for the coverage hole and the serious propagation attenuation in high-frequency bands \cite{ma2022cooperative}. However, these problem-specific algorithms are laborious processes and require much problem-specific knowledge, especially in multi-RISs aided systems.

Inspired by the recent successes of deep learning technology in many application domains, e.g., the computer vision \cite{goodfellow2016deep}, deep neural network (DNN)  models have been utilized for RIS-aided communication systems in recent years \cite{gao2020unsupervised,xu2021robust}. However, such DNN models may not be very suitable for the problems in wireless communications. This is because DNNs fail to capture the underlying topology of wireless communication networks, resulting in dramatic performance degradation when the network size becomes large.  Thus, in order to improve the scalability and generalization of the network, a long-standing idea is to incorporate the structures of the target task into the neural network architecture \cite{bengio2021machine}. 

From this perspective, the graph neural network (GNN) model has emerged as a new direction to exploit the graph topology of wireless communication networks \cite{shen2022graph}. There are some works applying the GNN for resource allocation in device-to-device (D2D) communication networks \cite{shen2020graph,zhang2021scalable}.  In these works, the service links and interference relations are modeled as nodes and edges, respectively. Based on these graph embedding features, the optimization problem on the weighted sum rate (WSR) can be solved efficiently.

In this paper, we leverage the advantage of the GNN to solve the beamforming design in a multi-RISs aided communication system. Unfortunately, the modeling method for above D2D systems is not suitable for our system. In D2D systems, the transceiver pair (including one transmitter and one receiver) is denoted as one node without interference inside, and the interference in the system only exists between different transceiver pairs. However, we consider the communication links between one transmitter (BS) and multiple receivers (users) in this paper, where the inter-user interference is included. Thus, a new modeling approach is needed to better map the multi-RISs aided system into the graph. 

Aside from the beamforming design, we consider a more practical and general problem in the multi-RISs scenario, assuming that each user only selects one of the RISs to enhance the links' service quality. In fact, selecting all RISs is not an efficient way to assist each user, because RISs are generally deployed in different positions. The distances between the user and some RISs may be so far that the signals reflected by the RISs are severely attenuated. For these reasons, the RISs association is introduced into this problem which needs to be jointly optimized with the cooperative beamforming at both the BS and RISs to obtain better performance. Thus, our main contributions in this work are concluded as 

1) We first propose a new GNN structure to map the optimization problem in a multi-RISs aided system and validate its effectiveness through extensive experiments. 

2) A RISs association strategy is considered in our optimization problem. By comparing the results obtained with fully connected RISs, we find that the strategy can efficiently improve the quality services of users  by  30$\%$.

3)In order to exhibit the benefits of applying structure information, DNNs with different network sizes are adopted as benchmarks. It is clear that our proposed heterogeneous GNNs perform about 10 times better than DNNs.

\section{Multi-RISs Aided Massive MIMO Communication Networks}
In this section, we introduce a multi-RISs aided MIMO system model with a RISs association strategy.

\subsection{System Model}
 \begin{figure}[h]
\setlength{\abovecaptionskip}{0pt}
\setlength{\belowcaptionskip}{0pt}
\centering
\includegraphics[width= 0.4\textwidth]{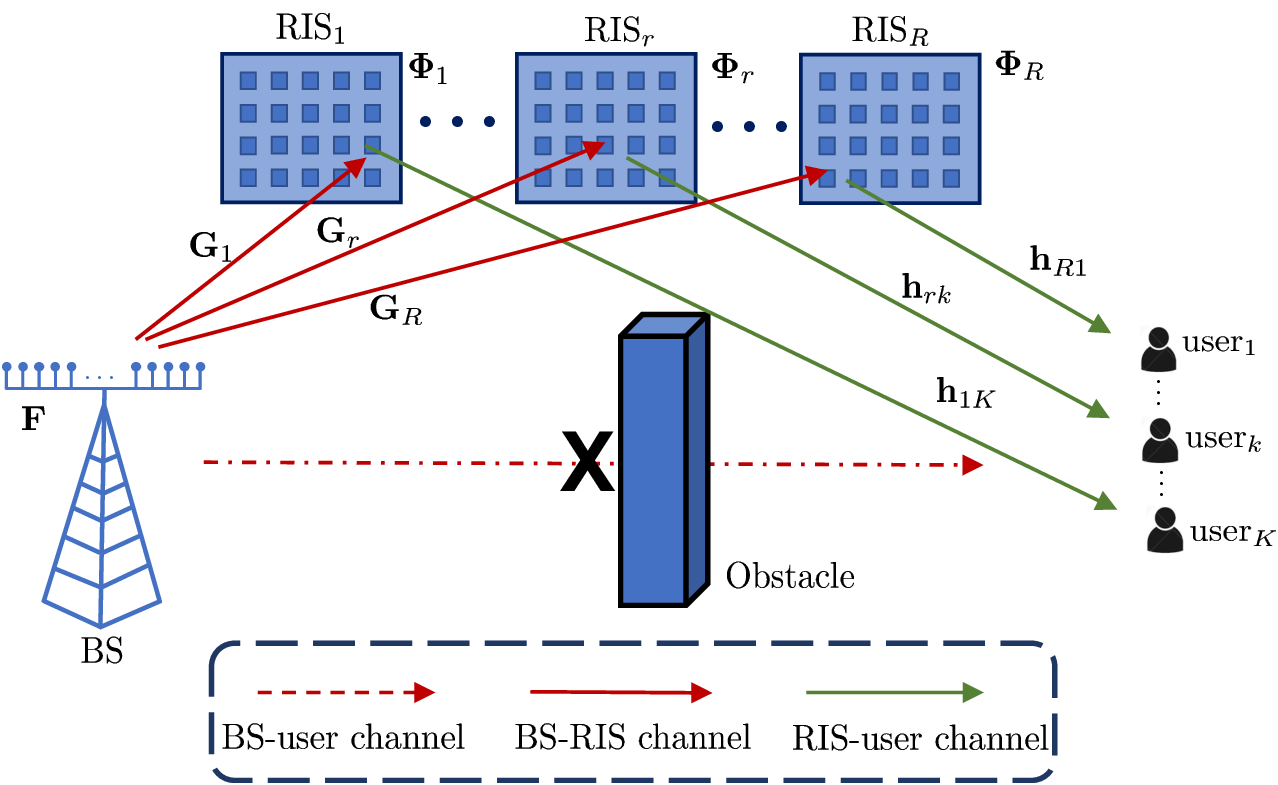}
\DeclareGraphicsExtensions.
%\captionsetup{font={scriptsize}}
\caption{Illustration of the multi-RISs aided massive MIMO system model.}
%\label{1}
\vspace{-5pt}
\end{figure}
We consider a multi-RISs aided system with  a $N_t$-antenna BS, a set of single-antenna users $\mathcal{K} = \left\{1,2,\cdots,K \right\} $ and a set of RISs $\mathcal{R} = \left\{1,2,\cdots, R \right\}$. Each RIS is equipped with $M = M_x \times M_y$ reflective elements, where $M_x$ and $M_y$ represent the the number of horizontal and vertical elements, respectively. Assume that the BS transmits $K$ data streams to users through the beamforming architecture. Thus, the transmitted signal $\mathbf{x}$ from the BS can be expressed as
\begin{align}
\mathbf{x} 
   &\triangleq \mathbf{F} \mathbf{s} 
   = \sum_{k = 1}^{K} \mathbf{f}_k {s}_{k},
\end{align}
where $\mathbf{s} = \left[ s_1,\cdots,s_K\right]^{T} \in \mathbb{C}^{K \times 1}$ indicates the symbol vector and $\mathbb{E}({\mathbf{s}\mathbf{s}^H}) = \mathbf{I}_K$.  $\mathbf{F} = \left[\mathbf{f}_{1}, \cdots, \mathbf{f}_{K} \right]  \in \mathbb{C}^{N_\mathrm{t} \times K}$ denotes the beamforming matrix at the BS and the power constraint is denoted by  $\left\|\mathbf{F}\right\|_\mathcal{F}^2 \leq P_{max}$, where $P_{max}$ is the maximum transmit power by the BS. It is crucial to investigate the joint design of RISs association, BS beamforming, and the phase configuration of RISs. In this paper, we choose one of all RISs for each user to maximize the WSR of the multi-RISs aided massive MIMO system. To simply express the association between RISs and users, a matrix $\mathbf{U} = [\mathbf{u}_1, \cdots, \mathbf{u}_K]^{T} \in \mathbb{C}^{K \times R}$ consisting of binary variables $u_{k,i}$ is defined as  
\begin{equation}
   u_{k,i}=\left\{\begin{array}{l}
1,  \text {user}_k \text { is assisted by RIS}_i. \\
0, \text { otherwise. }
\end{array}\right.
\end{equation}

Thus, based on the RISs association strategy, the equivalent channel $\mathbf{h}_{k}$ from the BS to user$_k$ can be expressed as
\begin{equation}
\mathbf{h}_{k} 
   =\sum_{i \in \mathcal{R}} u_{k, i} \mathbf{h}_{ik} {\mathbf{\Phi}}_{i}\mathbf{G}_{i},
\end{equation}
where $\mathbf{G}_{i} \in \mathbb{C}^{M \times N_t}$ and  $\mathbf{h}_{ik} \in \mathbb{C}^{1 \times M}$  denote the BS-RIS$_i$ and RIS$_i$-user$_k$, respectively. $\boldsymbol{\Phi}_{i} = {\rm{diag}} \left( \theta_{i,1},\cdots,\theta_{i,M} \right) \in \mathbb{C}^{M \times M}$ with the diagonal element $\theta_{i,m} = e^{j\phi_{i,m}} $ denoting the phase shift matrix of RIS$_i$. 
%Thus, the signal $\mathbf{y}_{k}$ received by the user$_k$ can be rewritten as
% \begin{align}
% \mathbf{y}_{k}
% &= \left( \sum_{i \in \mathcal{R}} u_{k,i} \mathbf{h}_{ik} {\mathbf{\Phi}}_{i}\mathbf{G}_{i} \right)\mathbf{f}_{k}{s}_{k}\\\nonumber
% &+\sum_{j\in \mathcal{K}, j \neq k}^{K} \left( \sum_{i \in \mathcal{R}} u_{k,i} \mathbf{h}_{ik} {\mathbf{\Phi}}_{i}\mathbf{G}_{i}  \right)  \mathbf{f}_{j} {s}_{j} +  \mathbf{n}_k.
% \end{align}

According to above assumption, the signal to interference plus noise ratio (SINR) at user$_k$ can be  given as
\begin{align}
{\rm{SINR}}_{k} = \frac{\left| \mathbf{h}_k  \mathbf{f}_{k}\right|^{2}}{ \sum_{j\in \mathcal{K}, j \neq k}^{K} \left|\mathbf{h}_k  \mathbf{f}_{j}
\right|^{2}+\sigma^{2}},
\end{align}
where $\sigma^{2}$ denotes the effective noise variance.
\subsection{Channel Model}

In this paper, we consider the mmWave channel and adopt the typical Saleh-Valenzuela channel model\cite{akdeniz2014millimeter} with limited scattering paths to characterize the sparse nature. Supposing a uniform linear array (ULA) is equipped at the BS and a uniform planar array (UPA) is equipped at each RIS. Hence, we can get the array response vectors by the following equations
\begin{align}
\mathbf{a}_{\mathrm{ULA}}(\phi) &= \frac{1}{\sqrt{N}}\left[e^{-j \frac{2 \pi d}{\lambda} \phi_i}\right]_{i \in \mathcal{I}(N)}, \\
 \mathbf{a}_{\mathrm{UPA}}(\phi^1,\phi^2) 
 &= \mathbf{a}_{\mathrm{ULA}}^{\mathrm{x}}(\phi^1) \otimes \mathbf{a}_{\mathrm{ULA}}^{\mathrm{y}}(\phi^2),
\end{align}
where $\lambda$ and $d$ represent the signal wavelength and the antenna spacing, respectively. As for a ULA with $N$ antennas, $\phi$ denotes the angle of arrival (AoA) or the angle of departure (AoD). For a UPA with $M = M_y \times M_x$ elements, $\phi^1$ and $\phi^2$ denote the azimuth and elevation angles, respectively. $\mathbf{a}_{\mathrm{ULA}}^{\mathrm{x}}(\phi^1)$ and $\mathbf{a}_{\mathrm{ULA}}^{\mathrm{y}}(\phi^2)$ are defined in the same manner as $\mathbf{a}_{\mathrm{ULA}}(\phi)$ with $\mathcal{I}(N)=\{n-(N-1) / 2, n=0,1, \cdots, N-1\}$, and $\otimes$ denotes the Kronecker product. The array element spacings of ULA and UPA are assumed to be $\lambda / 2$. Hence, the effective channels can be represented as 
%both the direct BS-user channels $\mathbf{D}_{k}$ and cascaded BS-RIS-user channels ${\mathbf{H}}_{r, k}$, $\mathbf{G}_{r} $ can be represented as
 \begin{align}
   \mathbf{h}_{ik} &= \sum_{l_h=1}^{L_h}\beta_{l_h} {\mathbf{a}}_{\mathrm{r}}^H \left(\phi_{l_h}^{{\mathrm{r}},1}, \phi_{l_h}^{{\mathrm{r}},2} \right), \\
    \mathbf{G}_{i} &= \sum_{l_g=1}^{L_g}\beta_{l_g}{\mathbf{a}}_{\mathrm{r}} \left(\phi_{l_g}^{{\mathrm{r}},1}, \phi_{l_g}^{{\mathrm{r}},2} \right) {\mathbf{a}}_{
    \mathrm{b}}^H \left(\phi_{l_g}^{\mathrm{b}} \right),
\end{align}
where $L_s$ is the number of paths that contains one LoS path and $(L_s-1)$ $\mathrm{NLoS}$ paths for link, and $\beta_{l_s}$ denotes the complex path gain of $l_{s}$ path with $ s \in \left\{h, g\right\}$. For simply notation, the subscript $h, g$ represent $\mathbf{h}_{ik}$ and  $\mathbf{G}_i$, respectively. 

\subsection{Problem Formulation}

Our main objective is to maximize the WSR by designing the beamforming matrix $\mathbf{F}$, the RISs association matrix $\mathbf{U}$ as well as the passive beamforming matrices $\mathbf{\Phi}_i, \forall i \in \mathcal{R}$ in multi-RISs aided communication systems. Therefore, the maximization problem can be formulated as
%. First, the WSR is given by
% \begin{align}
%     R  &= \sum_{k = 1}^{K} w_k\rm{log}_2(1+ {\rm{SINR}}_k),
% \end{align}
 %Subsequently, the maximization problem can be formulated as:
\begin{subequations}
\begin{align}
\text { P1: }
{\mathop{\max}\limits_{\mathbf{F}, \mathbf{U}, \mathbf{\Phi}_i,\forall i \in \mathcal{R} }} & \sum_{k = 1}^{K}w_k \rm{log}_2(1+ {\rm{SINR}}_k) \\
\text { s.t. } 
& \left\|\mathbf{F} \right\|_\mathcal{F}^{2} \leq P_{max} \\
& \left|\boldsymbol{\Phi}_{i,m}\right| = 1,  \forall m = 1,2, \cdots, M \\
& \sum_{i \in \mathcal{R}} u_{k,i} = 1, \forall k \in \mathcal{K} \\
& u_{k,i} = \left\{0,1\right\}, \forall k \in \mathcal{K}, \forall i \in \mathcal{R},
\label{P1}
\end{align}
\end{subequations}
where $w_k$ denotes the weight for user$_k$ with $\sum_{k = 1}^K w_k = 1$.

\section{Sum-Rate Maximization Via GNN}

%  \begin{figure}[h]
% \setlength{\abovecaptionskip}{0pt}
% \setlength{\belowcaptionskip}{0pt}
% \centering
% \includegraphics[width= 0.4\textwidth]{graph_representation.pdf}
% \DeclareGraphicsExtensions.
% %\captionsetup{font={scriptsize}}
% \caption{Illustration of the graph model in a multi-RISs aided system.}
% \label{2}
% \vspace{-5pt}
% \end{figure}

\begin{figure*}[ht]
\centering
\includegraphics[scale=.5]{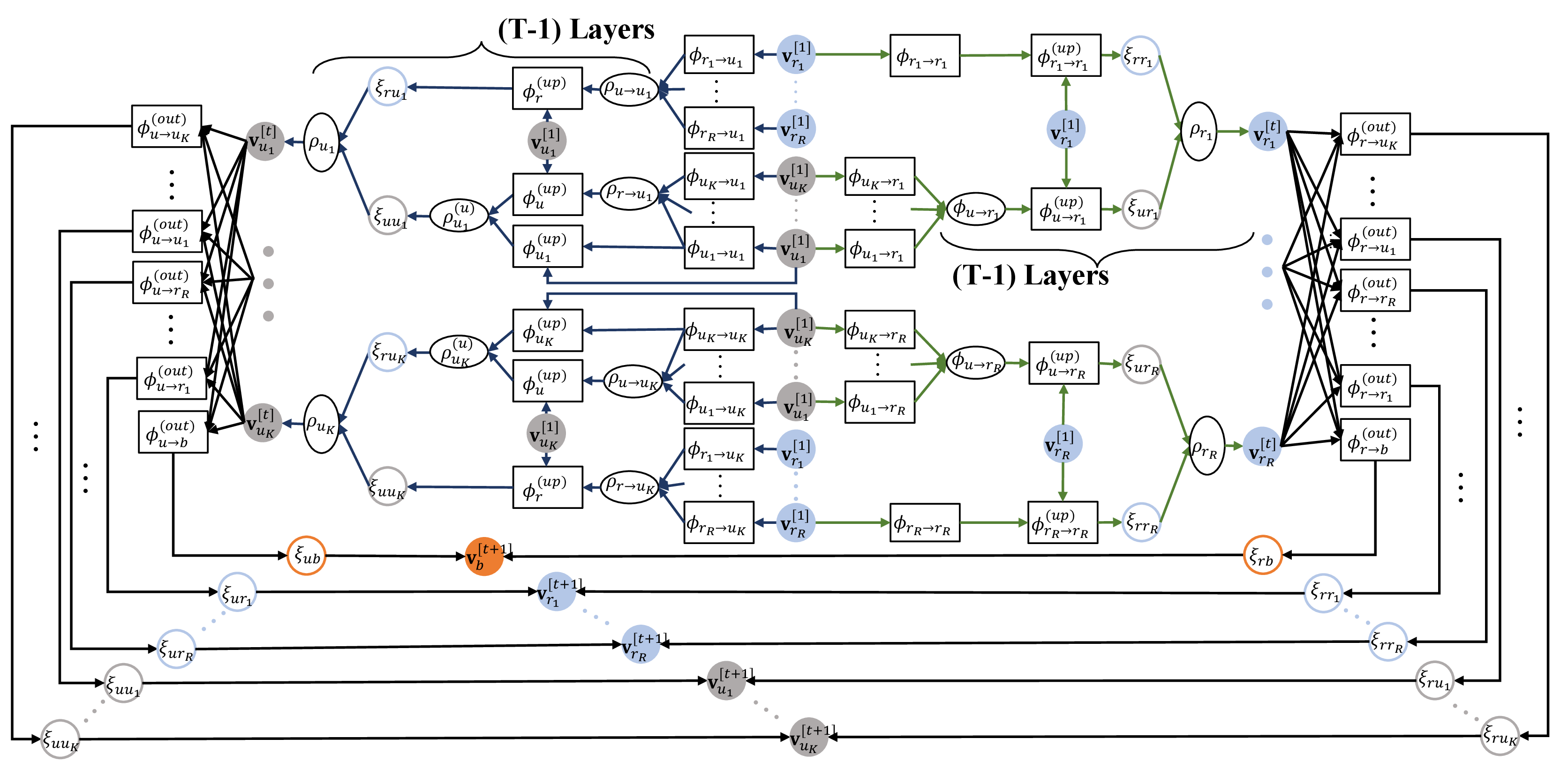}
\caption{The message passing paradigm in the heterogeneous GNN network.}
\label{3}
\end{figure*}
\vspace{-6pt}

 In this paper,we introduce the heterogeneous GNN which contains abundant information with structural relations among multi-typed nodes as well as unstructured content associated with each node \cite{zhang2019heterogeneous}. 
 As for a heterogeneous graph $\mathcal{G}=$ $(\mathcal{V}, \mathcal{E}, \mathcal{O}_\mathcal{V}, \mathcal{R}_\mathcal{E})$, there are two additional type sets, including the predefined node types $\mathcal{O}_\mathcal{V}$ and edge types $\mathcal{R}_\mathcal{E}$ with $|\mathcal{O}_\mathcal{V}|+|\mathcal{R}_\mathcal{E}|>2$ \cite{schlichtkrull2018modeling}.  
 The node types $\mathcal{O}_\mathcal{V}$ includes {BS node} $\mathbf{v}_{b}$, {RIS$_i$ node} $\mathbf{v}_{r_i}$ and {user$_k$ node} $\mathbf{v}_{u_k}$. According to the wireless communication scenario, the attributes of $\mathbf{v}_{u_k}$ include the weight $w_k$, noise variance $\sigma^2$ as well as $\mathbf{U}$. Similarly, the attributes $\mathbf{v}_{b}$ and $\mathbf{v}_{r_i}$ include $\mathbf{F}$ and  ${\mathbf{\Phi}}_{i}$, respectively.  Given the heterogeneous GNN structure,  the goal of interest is to find a policy  $p_\Theta(\cdot)$ with parameters $\Theta$, mapping the graph $\mathcal{G}$ to estimations of the optimal ${\mathbf{F},\mathbf{\Phi}_i}$ and $\mathbf{U}$.

 \subsection{The Initialization Layer of Proposed GNN}

 The initial layers take the CSI information as the input features. In order to better learn the phase configuration of each RIS, we extract the phase shift matrix $\boldsymbol{\theta}_i$. Thus, the equivalent channel $\mathbf{h}_k$ can be expressed as  
 \begin{align}
    \mathbf{h}_{k} 
   &= \sum_{i \in \mathcal{R}} u_{k, i} {\boldsymbol{\theta}}_i
   \mathrm{diag}(\mathbf{h}_{ik}) \mathbf{G}_{i} \\ \nonumber
   &= \sum_{i \in \mathcal{R}} u_{k, i} {\boldsymbol{\theta}}_i \mathbf{H}_{ik}^{\mathrm{cas}}, 
 \end{align}
where ${\boldsymbol{\theta}}_i = \left[\theta_{i,1},\cdots,\theta_{i,M}\right]$ and  $\mathbf{H}_{ik}^{\mathrm{cas}}= \mathrm{diag}(\mathbf{h}_{ik}) \mathbf{G}_{i}$ denotes the cascaded channel links, which is considered as the input feature for each user node, i.e., $\mathbf{v}_{u_k}^{(\mathrm{in})} = \left[ \operatorname{Re}(\mathbf{H}_{ik}^{\mathrm{cas}}), \operatorname{Im}(\mathbf{H}_{ik}^{\mathrm{cas}}) \right]$.

\subsection{Message Passing Paradigm}

Graph network block is adopted as the basic computation unit over graphs \cite{battaglia2018relational}, as illustrated in Fig. \ref{3}. The update function for each type of node is illustrated in detail as follows.
 
\subsubsection{Update for RIS Nodes}

The updated feature for the RIS nodes at step $t$ $\mathbf{v}_{r_i}^{[t]}$ is obtained by combining the features from the last state $\mathbf{v}_{r_i}^{[t-1]}$ and all the user nodes  $\mathbf{v}_{u_k}^{[t-1]}, \forall k \in \mathcal{K}$. 
Then, the messages from different type of nodes are given as 
\begin{align}
&&\boldsymbol{\xi}_{rr_i} =  \phi_{r_i \rightarrow r_i}^{({\rm{up}})}(  \boldsymbol{\xi}_{r_i}, \mathbf{v}_{r_i}^{[t-1]}),
\end{align}
\begin{align}
 \boldsymbol{\xi}_{ur_i} &= \phi_{u \rightarrow r_i}^{({\rm{up}})}( \rho_{u \rightarrow r_i}( \boldsymbol{\xi}_{u_1} ,\cdots, \boldsymbol{\xi}_{u_K}), \mathbf{v}_{r_i}^{[t-1]}),
\end{align}
% where 
 where $\boldsymbol{\xi}_{\rm{src}} = \phi_{{\rm{src}} \rightarrow r_i}(\mathbf{v}_{\rm{src}}^{[t-1]})$ denotes message from the source node and ${\rm{src}} \in \{r_i, u_1, \cdots, u_K\} $ for simple notation. Meanwhile, both $ \phi_{{\rm{src}} \rightarrow r_i}$ and $\phi_{ {\rm{src}}\rightarrow r_i}^{({\rm{up}})}$ are two-layer MLPs. Besides, $\rho_{u \rightarrow r_i} $ is the mean aggregation function for all the different user nodes. Then, in heterogeneous GNNs, the updated RIS nodes $\mathbf{v}_{r_i}^{[t]}$ aggregate the feature information from different-type neighbors, which is given as 
\begin{align}
  \mathbf{v}_{r_i}^{[t]} = \rho_{r_i}( \boldsymbol{\xi}_{rr_i},  \boldsymbol{\xi}_{ur_i}) + \mathbf{v}_{r_i}^{[t-1]},  
\end{align}
where $\rho_{r_i}$ denotes the operation of the mean aggregation. Besides, a residual operation is added in this step to avoid degradation and gradient explosion of our network   \cite{2016Deep}.

 \subsubsection{Update for User Nodes}
 
The updated feature $\mathbf{v}_{u_k}^{[t]}$ is obtained by combining the features from all the user nodes $\mathbf{v}_{u_k}^{[t-1]}$  and RIS nodes $\mathbf{v}_{r_i}^{[t-1]}$. In the same manner, the messages from different type of nodes are given as 
\begin{align}
 {\boldsymbol{\xi}}_{ru_k} &= \phi_{r \rightarrow u_k}^{({\rm{up}})}( \rho_{r \rightarrow u_k}( \boldsymbol{\xi}_{r_1} ,\cdots, \boldsymbol{\xi}_{r_R}), \mathbf{v}_{u_k}^{[t-1]}),
\end{align}
\begin{align}
 {\boldsymbol{\xi}}_{uu_k} = \rho_{u_k}^{(u)} ( \phi_{u \rightarrow u_k}^{({\rm{up}})} (\boldsymbol{\xi}_{u}),  \phi_{u_k\rightarrow u_k}^{(up)}(\boldsymbol{\xi}_{u_k},
 \mathbf{v}_{u_k}^{[t-1]})) ),
\end{align}
where $\boldsymbol{\xi}_{u} = \rho_{u \rightarrow u_k}( \boldsymbol{\xi}_{u_1} ,\cdots, \boldsymbol{\xi}_{u_K})$ and $\boldsymbol{\xi}_{\rm{src}} = \phi_{{\rm{src}} \rightarrow r_i}(\mathbf{v}_{\rm{src}}^{[t-1]})$ denotes message from the source node and ${\rm{src}} \in \{r_1, \cdots, r_R, u_1, \cdots, u_K\} $ for simple notation. Meanwhile, both $ \phi_{{\rm{src}} \rightarrow u_k}$ and $\phi_{ {\rm{src}}\rightarrow u_k}^{({\rm{up}})}$ are two-layer MLPs. Besides, $\rho_{u \rightarrow u_k}$ and $\rho_{r \rightarrow u_k}$ are the max and mean aggregation functions, respectively. Then, in heterogeneous GNNs, the updated RIS nodes $\mathbf{v}_{u_k}^{[t]}$ aggregate the feature information from different-type neighbors, which can be obtained by
\begin{align}
  \mathbf{v}_{u_k}^{[t]} = \rho_{u_k}( \boldsymbol{\xi}_{ru_k},  \boldsymbol{\xi}_{uu_k}) + {\mathbf{v}}_{u_k}^{[t-1]},  
\end{align}
where $\rho_{u_k}$ denotes the operation of the max aggregation.
 
\subsubsection{Output Layers for Node Features}

After updating all the nodes, each node has multiple feature embeddings from different-type nodes. In order to obtain the final feature of each node $\mathbf{v}_{b}^{[T]}$, $\mathbf{v}_{r_i}^{[T]}$ and $\mathbf{v}_{u_k}^{[T]}$, a decoder layer is used to output all the optimized features, which is given as
\begin{align}
  \mathbf{v}_{\rm{dst}}^{[T]} = \rho_{ \rm{dst}}^{\rm{(out)}}( \boldsymbol{\xi}_{u{\rm{dst}}}, \boldsymbol{\xi}_{r{\rm{dst}}}),  
\end{align}
where the destination node is denoted as ${\rm{dst}} \in \{b, r_1, \cdots, r_R, u_1, \cdots, u_K\}$ and  $\phi_{\rm{src} \rightarrow \rm{dst}}^{(\rm{out})}$ is utilized to extract the features from different-type nodes, as shown by the black line in Fig. \ref{3} . Finally, the new state $\mathbf{v}_{\rm{{{dst}}}}^{[T]}$ will be obtained by a mean aggregation layer $\rho_{\rm{dst}}^{(\rm{out})}$.

 Suppose that $T$ iterations are executed during the training process, and we defined each update as HGN$_t, \forall t \in T$. An encoder-process-decoder architecture is utilized to facilitate training \cite{battaglia2018relational}, including one encoder block HGN$_0$, one decoder block HGN$_T$, and one core block (concatenate by HGN$_1,\cdots$, HGN$_{T-1}$). After the encoder-process-decoder architecture, we can obtain the final result for the optimal $\mathbf{F}$, $\mathbf{U}$ and $\mathbf{\Phi}_i, \forall i \in \mathcal{R}$.

\subsection{Model Architecture of Proposed Heterogeneous GNN}

In addition to the massage passing paradigm, the proposed heterogeneous GNN needs some additional layers to complete the whole learning process as follows

\subsubsection{The Normalization Layers} Note that the neural network is based on the real numbers, 
%we cannot directly learn the mapping of the channel information to the optimization values, which are all complex values in wireless communication systems. Instead,
we learn the vector consisting of the real and imaginary parts of the beamforming matrix ${\mathbf{F}}$ and the phase matrices $\mathbf{\Phi}_i, \forall i \in \mathcal{R}$. Therefore, before calculating the WSR loss function, additional layers are necessary to merge the real-valued results to complex-valued results. 

\begin{itemize}
\item To satisfy the constraint on transmit power in (9b), a normalization layer is employed at the output feature $\operatorname{Re} \{{\mathbf{F}} \}$and $\operatorname{Im} \{{\mathbf{F}} \}$, which can be expressed as
\begin{align}
 \hat{\mathbf{F}}=\sqrt{{P}_{max}} \frac{\operatorname{Re}\{{\mathbf{F}}\}+j \cdot \operatorname{Im}\{\mathbf{F}\}}{\sqrt{\left\|\operatorname{Re}\{\mathbf{F}\}\right\|_{\mathcal{F}}^2+\left\|\operatorname{Im}\{\mathbf{F}\}\right\|_{\mathcal{F}}^2}}.   
\end{align}
 
\item Similarly, the normalization layer for the reflection coefficients can be represented as
% the phase configuration on each RIS has the unit-modulus constraint (9c), and a normalization layer is utilized to tackle the real and imaginary components of the reflection coefficients, which can be represented as
\begin{align}
 {{\theta}}_{i,m} = \frac{\operatorname{Re} \{ {{\theta}}_{i,m} \}+ j \cdot \operatorname{Im} \{ {{\theta}}_{i,m} \} }{ \sqrt{ |\operatorname{Re} \{  {{\theta}}_{i,m} \}|^2 + |\operatorname{Im} \{  {{\theta}}_{i,m} \} |^2 } }.
\end{align} 
\end{itemize}

\subsubsection{The RIS Association Strategy} Each user chooses a suitable one of all RISs as the reflection link to enhance the link quality, as the constraint (9d).  Thus, the matrix $\mathbf{U}$ for calculating the WSR can be obtained by 
\begin{align}
    c_{u_{k,i}} = \operatorname{SoftMax}\left(u_{k,i}\right)=\frac{e^{u_{k,i}}}{\sum_{i=1}^R e^{u_{k,i}}}.
\end{align}

After obtaining the SoftMax score $c_{u_{k,i}}$, we sort the scores of each user, i.e., each column in $\mathbf{U}$. We select the largest value and set it to 1, indicating that the user$_k$ selects this RIS to assist the communication links while setting the remaining positions of this column to 0 to get the final matrix. 

\subsubsection{The Loss Function}  Due to the connection matrix is composed of binary elements, which are discontinuous variants. Thus, the process to determine $\mathbf{U}$ as well as the optimization of $\mathbf{F}$ and $\mathbf{\Phi}$ is very difficult to learn. 
% In order to better learn the RIS association matrix,  it is important to have a suitable optimization criterion.
Considering this, we add a regular term related to the distance between users and RISs to the training loss. We adopt distance-dependent labels to represent the association relationship between the users and their nearest RIS and the categorical cross-entropy loss function to achieve it.  Thus, the loss function is composed of the negative WSR and the categorical cross-entropy loss function,  which can be expressed as
\begin{align}
 &\operatorname{Loss} =  \\\nonumber
 &- \sum_{k = 1}^{K}\left( w_{k}\rm{log}_2\left(1 \!+\! \frac{\left| \mathbf{h}_{\it{k}} \mathbf{f}_{\it{k}}\right|^{2}}{ \sum\limits_{ \it{j}\in \mathcal{K}, \it{j} \neq \it{k}}^{\it{K}} \left|
\mathbf{h}_{\it{k}} \mathbf{f}_{\it{j}}
\right|^{2}\!+\!\sigma^{2}} \right) \!-\!\eta \mathbf{u}_{k}^{(l)}  \log \mathbf{u}_k \right),
 \label{loss}
\end{align}
where ${\mathbf{u}}_{k}^{(\mathrm{l})}$ and $\mathbf{u}_k$ denote the distance-dependent label and optimized value for the user$_k$, respectively.  $\eta$ is the coefficient of the penalty term, and its role is to strike the balance of the significance of these two parts in the loss function, which can be calculated by
$\eta ={{\rm{WSR}}_p}/{{\rm{WSR}}_{p_0}}$. We set ${{\rm{WSR}}_{p}}$ as the results in the pre-training process where the loss function is negative WSR, and ${{\rm{WSR}}_{p_0}}$ is the baseline case when $P_{max} = 30 $dBm. For the training details, Adam optimizer \cite{kingma2014adam} is adopted with a total of 200 epochs on one NVIDIA RTX 2080Ti GPU. The initial learning rate, weight decay, and batch size are set to $1 \times 10^{-4}$, $5 \times 10^{-5}$ and 128.

\section{Simulation Results}

In this section, numerical simulations are provided to verify the effectiveness of our proposed heterogeneous GNN. Throughout the simulation, the BS is equipped with $N_t = 8$ antennas, and the number of RIS elements $M$ is 16 with $M_x = M_y = 4$. The BS is assumed to be located on the origin and two RISs are located in $[30,25]$ and $[30,-25]$. In addition, two users are randomly located in a rectangular region in $[40:50,-25:25]$. Without loss of generality, the weights $w_k$ are set to be equal. According to \cite{akdeniz2014millimeter}, the channel gain is taken as $\beta \sim \mathcal{C N}\left(0,10^{-0.1 \mathrm{PL}(r)}\right)$ where $\mathrm{PL}(r)= \varrho_a + 10 \varrho_b lg(r) + \xi$ with $\xi \sim \mathcal{N}(0,\sigma_{\xi}^2)$. In addition, the channel realizations are produced by setting $\sigma_n ^2 = -85 dBm$, $\varrho_{\mathrm{a}}=61.4, \varrho_{\mathrm{b}}=2$, and $\sigma_{\xi}=5.8 \mathrm{~dB}$. The total number of samples is 110,000 where 100,000 and 10,000 of the whole generated data are used for training and validation, respectively. To see the performance, WMMSE with a random phase configuration is used as the performance baseline. Also, in order to validate the effectiveness of the graph structure, we compare the performance in our proposed heterogeneous GNN with that in DNN.
 \begin{figure}[h]
\setlength{\abovecaptionskip}{0pt}
\setlength{\belowcaptionskip}{0pt}
\centering
\includegraphics[width= 0.45\textwidth]{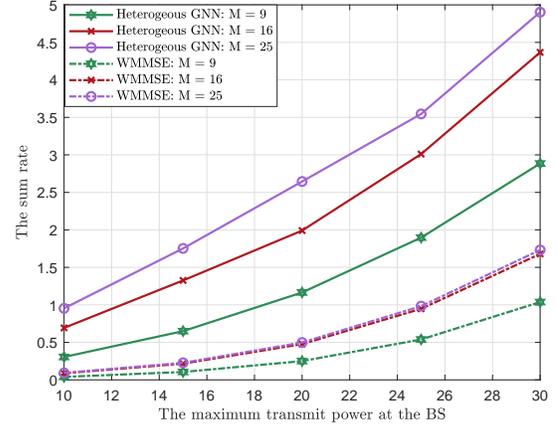}
\DeclareGraphicsExtensions.
%\captionsetup{font={scriptsize}}
\caption{The sum rate versus the maximum transmit power at the BS.}
\label{6}
\vspace{-5pt}
\end{figure}

 Fig. \ref{6} shows the sum rate with respect to the maximum transmit power of the BS. Our proposed GNN structure performs well under all levels of transmit power, which exhibits the generality and effectiveness of our structure. First, the sum rate obtained by random configuration and heterogeneous GNN becomes higher with the increase of maximum transmit power $P_{max}$. However, in the absence of optimization, the sum rate boosting from the increasing power is limited. In contrast, our proposed GNN structure exhibits its superb ability to solve the optimization problem P1 with different number of reflective elements (M = 9, 16, 25) and different $P_{max}$, from $10$ dBm to $30$ dBm with an interval of $5$ dBm. 
 \begin{figure}[h]
\setlength{\abovecaptionskip}{0pt}
\setlength{\belowcaptionskip}{0pt}
\centering
\includegraphics[width= 0.45\textwidth]{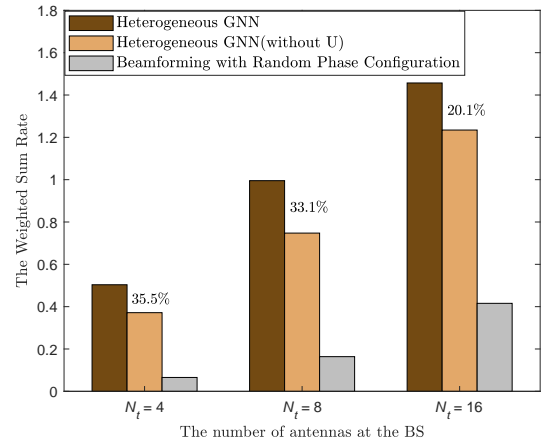}
\DeclareGraphicsExtensions.
%\captionsetup{font={scriptsize}}
\caption{The weighted sum rate versus the number of antennas at the BS}
\label{5}
\vspace{-5pt}
\end{figure}

 Fig. \ref{5} illustrates the total number of antennas with respect to the WSR performance. We also exhibit the WSR values in two other cases, including the results by WMMSE with the random phase configuration and the optimal value obtained by our proposed GNN in the case of one fixed RIS, i.e., all the users receive the signal reflected by the RIS located at $[30,25]$. It is obvious that the WSR increase with the growth of the number of antennas under all cases, and heterogeneous GNN can achieve a higher WSR. %Meanwhile, by comparing the results of the fixed RIS, simulation results verify the necessity of the RISs association strategy. 
 We can choose a suitable one of all RISs to replace the fix RISs to provide service links for each user, which can improve the performance about $30\%$ under all the different number of antennas.
 \begin{figure}[h]
\setlength{\abovecaptionskip}{0pt}
\setlength{\belowcaptionskip}{0pt}
\centering
\includegraphics[width= 0.45\textwidth]{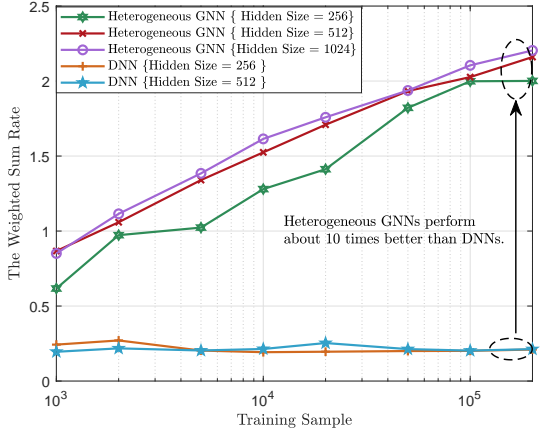}
\DeclareGraphicsExtensions.
%\captionsetup{font={scriptsize}}
\caption{The weighted sum rate versus the number of training sample.}
\label{4}
\vspace{-5pt}
\end{figure}

 Fig. \ref{4} illustrates the WSR of heterogeneous GNNs and two DNNs with respect to the same size of the training set under the networks with different sizes. We can easily find that the performance in heterogeneous GNNs performs significantly better  than those of DNNs, especially when the number of training sample is $2 \times 10^{5}$, GNNs perform 10 times better than DNNs. Since the introduction of the RISs association strategy will exacerbate the influence of structural information on optimization results, DNNs perform very poorly due to their lack of the ability to capture the underlying model structure. In addition, comparing the results of two DNNs, it can be seen that no matter how to increase the number of layers and the number of channels cannot improve the performance of DNNs.  In addition, GNNs can get better optimization results with fewer samples, even if it is less than 10,000, and get better results with increasing training samples.
 
\section{Conclusion}

We proposed a heterogeneous GNN structure to solve the NP-hard beamforming optimization problem in a multi-RISs aided MIMO system by mapping the structure of the wireless system into a graph neural network. Different from most beamforming problems, a RISs association strategy was introduced into our problem, which makes the beamforming design more complex, but a suitable RISs association strategy can enhance the users’ quality of service. The effectiveness of our proposed GNN structure and the RISs association strategy were validated through extensive simulations. According to the simulation results, we found that our proposed  GNNs performed about 10 times better than the conventional DNNs. Moreover, the benefits of the RISs association strategy were illustrated by comparing the optimal results with/without this strategy, which improved the performance about $30\%$. 

\bibliographystyle{IEEEtran}
\bibliography{IEEEabrv,myref}

\end{document}